\documentclass[aps,prd,amsmath,showpacs,nofootinbib]{revtex4}
\usepackage[active]{srcltx}
\usepackage{epsfig,amsmath,amsfonts,wasysym}
\usepackage{dsfont,graphicx}

\newcommand{\Ref}[1]{(\ref{#1})}
\newcommand{\be}{\begin{equation}}
\newcommand{\ee}{\end{equation}}
\newcommand{\beq}{\begin{equation}}
\newcommand{\eeq}{\end{equation}}
\newcommand{\bn}{\begin{eqnarray}}
\newcommand{\en}{\end{eqnarray}}
\newcommand{\bd}{\begin{displaymath} }
\newcommand{\ed}{\end{displaymath}}
\newcommand{\bnn}{\begin{eqnarray*}}
\newcommand{\enn}{\end{eqnarray*}}
\newcommand{\bs}{\begin{subequations}}
\newcommand{\es}{\end{subequations}}
\newcommand{\adb}{\allowdisplaybreaks }

\newcommand{\phisq}{\langle\phi^2\rangle}
\newcommand{\Tmn}{\langle T_{\mu\nu}\rangle}

\begin{document}
\title{Vacuum stress-energy tensor of a massive scalar field in a wormhole spacetime}

\author{V.B. Bezerra$^1$}%
\email{valdir@fisica.ufpb.br}

\author{E.R. Bezerra de Mello$^1$}%
\email{emello@fisica.ufpb.br}

\author{N.R. Khusnutdinov$^2$}%
\email{nail@fisica.ufpb.br}

\author{S.V. Sushkov$^{2,3}$}%
\email{sergey_sushkov@mail.ru}%

\affiliation{$^{1}$Departamento de F\'{\i}sica, Universidade Federal da
Para\'{\i}ba, Caixa Postal 5008, CEP 58051-970 Jo\~ao Pessoa, Pb,
Brazil}

\affiliation{$^{2}$Department of Physics,
Kazan State University, Kremlevskaya str. 18, Kazan 420008,
Russia}

\affiliation{$^{3}$Department of Mathematics, Tatar State University of
Humanities and Education, Tatarstan str. 2, Kazan 420021, Russia}

\begin{abstract}
The vacuum average value of the stress-energy tensor of a massive
scalar field with non-minimal coupling $\xi$ to the curvature on
the short-throat flat-space wormhole background is calculated. The
final analysis is made numerically. It was shown that the
energy-momentum tensor does not violate the null energy condition
near the throat. Therefore, the vacuum polarization cannot
self-consistently support the wormhole.
\end{abstract}

\maketitle

\section{Introduction}
Traversable wormholes have been introduced into physical realm by
Morris and Thorne \cite{MT} in 1988. At the same time they
realized that a matter threading a wormhole's throat should
possess {\em exotic}\, properties, namely it should have a
negative pressure and violate the null energy condition (NEC).
Later on, this result was generalized for {\em any} traversable
wormholes, both static and non-static \cite{HocVis}. As is known,
the classical matter does satisfy the usual energy conditions,
hence traversable wormholes cannot arise as solutions of classical
relativity and matter. Already in 1989 Morris, Thorne, and
Yurtsever \cite{MTY} supposed that quantized fields could play a
role of the exotic matter maintaining wormholes. Their reasons
were founded on the important fact stating that quantum field
theory may have states with negative energy density, thus
violating the weak energy condition \cite{EpsGlaJaf} (see, also,
\cite{Kuo}).

In the absence of the complete theory of quantum gravity, the
semi-classical approach gives the more natural way to include
quantized fields in the theory of gravity. Various wormhole
solutions in semi-classical gravity have been considered in the
literature. For instance, semi-classical wormholes were found in
the framework of the Frolov-Zelnikov approximation for $\langle
T_{\mu\nu}\rangle^{\rm ren}$ \cite{Sus92}. Analytical
approximations of the stress-energy tensor of quantized fields in
static and spherically symmetric wormhole space-times were also
explored in Refs. \cite{PopSus01,Pop01}. Some arguments in favor
of the possibility of existence of semi-classical wormholes have
been given by Khatsymovsky \cite{Kha}. However, the first
self-consistent wormhole solution coupled to a quantum scalar
field was obtained in Ref. \cite{HocPopSus}. The ground state of a
massive scalar field with a non-conformal coupling on a
short-throat flat-space wormhole background was computed in Ref.
\cite{KhuSus02}, by using a zeta renormalization approach. The
latter wormhole model, which was further used in the context of
the Casimir effect \cite{KhaKhuSus}, was constructed by excising
spherical regions from two identical copies of Minkowski
space-time, and finally surgically grafting the boundaries (A more
realistic geometry was considered in Ref. \cite{Khu03}). In a
series of works \cite{Gar05, Gar07, GarLob07, GarLob09} various
aspects of the graviton one loop contribution to a classical
energy in a wormhole background have been analyzed. The latter
contribution was evaluated through a variational approach with
Gaussian trial wave functionals, and the divergences were treated
with a zeta function regularization. In particular, the finite one
loop energy was considered as a self-consistent source for a
traversable wormhole.

Note that up to now no one has succeeded in exact calculations of
vacuum expectation values of the stress-energy tensor of quantized
fields on the wormhole background. The reason for this state of
affairs consists in considerable mathematical difficulties which
one faces with trying to quantize a physical field on the wormhole
background. To overcome these difficulties, in this work we will
consider a simple model of wormhole space-time given in
\cite{KhuSus02}: the short-throat flat-space wormhole. The model
represents two identical copies of Minkowski space with excised
from each copy spherical regions, and with boundaries of those
regions are to be identified. The space-time of this model is
everywhere flat except a two-dimensional singular spherical
surface. Due to this fact it turns out to be possible to construct
the complete set of wave modes of the massive scalar field and
calculate the stress energy tensor.

The organization of the paper is as follows. In Sec. \ref{Sec2} we
describe a space-time of wormhole in the short-throat flat-space
approximation. In Sec. \ref{Sec3} we analyze the solution of
equation of motion for massive scalar field and obtain Euclidean
Green function. Sec. \ref{Sec4} devoted to analysis of the vacuum
expectation value of the square of field from analytical and
numerical point of views. In Sec. \ref{Sec5} we obtain close
formulas for numerical analysis the stress-energy tensor and in
Sec. \ref{Sec6} discuss our results.

We use units $\hbar = c = G = 1$. The signature of the space-time,
the sign of the Riemann and Ricci tensors, is the same as in the
book by Hawking and Ellis \cite{HawEll73}.

\section{A traversable wormhole: the short-throat flat-space approximation}
\label{Sec2}
In this section we briefly consider a simple model of a
traversable wormhole (see Ref. \cite{KhuSus02}). Assume that the
throat of the wormhole is very short, and that curvature in the
regions outside the mouth of the wormhole is relatively weak. An
idealized model of such a wormhole can be constructed in the
following manner: Consider two copies of Minkowski space, ${\cal
M}_+$ and ${\cal M}_-$, with the spherical coordinates
$(t,r_{\pm},\theta_{\pm},\varphi_{\pm})$ [Notice: ${\cal M}_+$ and
${\cal M}_-$ have a common time coordinate $t$. One may interpret
this fact as the identification $t_+\leftrightarrow t_-$.]; excise
from each copy the spherical region $r_{\pm}<a$, where $a$ is a
radius of sphere; and then identify the boundaries of those
regions:
$(t,a,\theta_+,\varphi_+)\leftrightarrow(t,a,\theta_-,\varphi_-)$.
The Riemann tensor for this model is identically zero everywhere
except at the wormhole mouths where the identification procedure
takes place. Generically, there will be an infinitesimally thin
layer of exotic matter present at the mouth of the wormhole.

Such an idealized geometry can be described by the following
metric
\beq\label{metric}
ds^2=-dt^2+d\rho^2+r^2(\rho)\,(d\theta^2+\sin^2\theta\,d\varphi^2
),
\eeq
where $|\rho|$ is a proper radial distance, $-\infty<\rho<\infty$,
and the shape function $r(\rho)$ is
\beq
r(\rho)=|\rho\,|+a.
\eeq
It is easily to see that in two regions ${\cal R}_+\!:\,\rho>0$
and ${\cal R}_- \!:\,\rho<0$ separately, one can introduce a new
radial coordinate $r_\pm=\pm\rho+a$ and rewrite the metric
\Ref{metric} in the usual spherical coordinates:
\beq\label{Mink}
ds^2=-dt^2+dr_\pm^2+r_\pm^2(d\theta^2+\sin^2\theta\,d\varphi^2 ),
\eeq
This form of the metric explicitly indicates that the regions
${\cal R}_+\!:\,\rho>0$ and ${\cal R}_-\!:\,\rho<0$ are flat.
However, note that such the change of coordinates $r=|\rho\,|+a$
is not global, because it is ill defined at the throat $\rho=0$.
Hence, as was expected, the space-time is curved at the wormhole
throat. To illustrate this we explicitly calculate the scalar
curvature $R(\rho)$ in the metric \Ref{metric}:
\beq
R(\rho)=-8a^{-1}\delta(\rho)\label{ScaCur}.
\eeq

\section{Green function}\label{Sec3}

Let us discuss a massive scalar field $\phi$ with non-minimal
coupling to curvature. The scalar field equation of motion has the
following form
\beq
 (\square -\xi R - m^2)\phi = 0.
\eeq
A corresponding equation for the Euclidean Green function reads
\begin{equation}
 (\square_\tau -m^2 -\xi R)G_E(x;\tilde x) =
 -\frac{\delta^{(4)}(x,\tilde x)}{\sqrt{g}},
\end{equation}
where $\tau=-it$ is the Euclidean time. Due to spherical symmetry
one can represent $G_E(x;\tilde x)$ as follows
\begin{equation}\label{GE}
 G(x;\tilde x) = \sum_{l=0}^\infty \sum_{m=-l}^{+l} Y^*_{l,m}(\Omega)
Y_{l,m}(\Omega) \int \frac{d\omega}{2\pi} e^{i\omega\triangle\tau}
g_{\omega l}(\rho,\tilde\rho).
\end{equation}
Here $g_{\omega l}(\rho,\tilde\rho)$ is a radial Green function
obeying the equation
\begin{equation}\label{radialgreen}
 g_{\omega l}'' + \frac{2r'}{r}g_{\omega l}' - \left(\lambda^2+\frac{l(l+1)}{r^2} + \xi
R\right)g_{\omega l} = -\frac{\delta(\rho-\tilde\rho)}{r^2}.
\end{equation}
where $\lambda^2=\omega^2+m^2$ and a prime means the derivative
with respect to $\rho$. A solution of Eq. \Ref{radialgreen} can be
represented as
\beq
g_{\omega l}(\rho,\tilde\rho)=
\theta(\rho - \tilde\rho)\phi_+(\rho)\phi_-(\tilde\rho)+
\theta(\tilde\rho - \rho)\phi_+(\tilde\rho)\phi_-(\rho),
\eeq
where $\phi_\pm(\rho)$ are two independent solutions of the radial
equation of motion:
\begin{equation}\label{radphi}
\phi'' + \frac{2r'}{r}\phi' -
\left(\lambda^2+\frac{l(l+1)}{r^2}+\xi R\right)\phi = 0.
\end{equation}
In general, Eq. \Ref{radphi} cannot be solved in an explicit form.
However, this is possible in the particular case
$r(\rho)=|\rho|+a$. In this case $R = -8\delta (\rho)/a$ and Eq.
\Ref{radphi} yields
\begin{equation}\label{radphi1}
 \phi'' + \frac{2}{\rho \pm a}\phi' - \left(\lambda^2 + \frac{l(l+1)}{(\rho
\pm a)^2} - \frac{8\xi}{a}\delta(\rho)\right)\phi =0.
\end{equation}
Note that Eq. \Ref{radphi1} is an ordinary second-order
differential equation with a delta-like coefficient,
$\delta(\rho)$. A common treatment of such equations is to solve
them separately in regions with $\rho\not=0$ and then match the
obtained solutions and their first derivatives at $\rho=0$. In the
regions $\rho\not=0$ a general solution of Eq. \Ref{radphi1} reads
\beq\label{gensol}
\phi(\rho)= \sqrt{\frac{\pi}{2 r}}\,\left[C_1 I_\nu(\lambda r)+C_2
K_\nu(\lambda r)\right],
\eeq
where $C_1$ and $C_2$ are constants of integration and $\nu =
l+1/2$. Integrating Eq. \Ref{radphi1} around $\rho = 0$ gives the
following matching conditions at the throat
\begin{eqnarray}
 \phi(+0) - \phi(-0) &=& 0,\adb\nonumber\\
 \phi'(+0) - \phi'(-0) &=& -\frac{8\xi}{a} \phi(+0).\label{boundarycond}
\end{eqnarray}
Using the formula \Ref{gensol} and the relations
\Ref{boundarycond} one may define constants of integration $C_1$
and $C_2$ and then construct the radial Green function
$g(\rho,\tilde\rho)$ given by Eq. \Ref{radialgreen}. As the
result, one obtains (see for details Ref. \cite{BezKhu})
\begin{equation}\label{gr}
 g_{\omega l}(\rho,\tilde\rho) = g^{M}_{\omega l}(\rho,\tilde\rho)
 - \left.\frac{\lambda a(I_\nu K_\nu' +
    I_\nu'K_\nu)+(8\xi-1) I_\nu K_\nu}{2\lambda a K_\nu K_\nu'+
    (8\xi-1)K_\nu^2}\right|_{\lambda a}
    \frac{K_\nu(\lambda r)K_\nu(\lambda\tilde r)}{\sqrt{r\tilde
    r}},
\eeq
if $\rho$ are $\tilde\rho$ have the same signs, and
\beq
    g_{\omega l}(\rho,\tilde\rho)=\left.-\frac{1}
    {2\lambda a K_\nu K_\nu' + (8\xi-1)K_\nu^2}\right|_{\lambda a}
    \frac{K_\nu(\lambda r)K_\nu(\lambda\tilde r)}{\sqrt{r\tilde
    r}},
\end{equation}
if $\rho$ and $\tilde\rho$ have different signs. Here
\beq
g^M_{\omega l}(\rho,\tilde\rho)= \frac{K_\nu(\lambda
r)I_\nu(\lambda\tilde r)}  {\sqrt{r\tilde r}}
\eeq
is the radial Green function of the Minkowski spacetime.

\section{Vacuum polarization $\langle\phi^2\rangle$}\label{Sec4}
A vacuum of any quantized physical field is polarized in curved
space-time. A vacuum polarization $\phisq$ of the massive scalar
field $\phi$ in a wormhole space-time has been discussed in Refs.
\cite{Sus,PopSus01} in the WKB approximation. In this section we
represent exact calculations for $\phisq$ in a wormhole space-time
with the metric \Ref{metric}. The renormalized expression for the
vacuum polarization is defined as follows
\beq\label{phisq}
\phisq=\lim_{\tilde x\to x}\left[G_E(x;\tilde x)-G_{DS}(x;\tilde
x)\right],
\eeq
where $G_{DS}(x;\tilde x)$ are the well-known DeWitt-Schwinger
counterterms \cite{Chr78}:
\begin{eqnarray}
G_{DS}(x;\tilde x) &=& \frac{\triangle^{1/2}}{8\pi^2} \left\{ a_0
\left[ \frac{1}{\sigma} + m^2 L \left(1+ \frac 14 m^2 \sigma +
\cdots \right) - \frac 12 m^2 - \frac 5{16} m^4 \sigma + \cdots
\right]\nonumber \adb \right.\\
&-& \left. a_1 \left[L \left(1+ \frac 12 m^2 \sigma +
\cdots\right)  - \frac 12 m^2 \sigma - \cdots\right] + a_2\sigma
\left[ L \left( \frac 12 + \frac 18 m^2 \sigma + \cdots\right) -
\frac 14 - \cdots\right] + \cdots \adb\nonumber \right.\\ &+&
\left. \frac 1{m^2} [a_2 + \cdots] + \frac 1 {2m^4} [a_3 + \cdots]
+ \cdots \right\},
\end{eqnarray}
where $\sigma$ is half the squared distance between the points $x$
and $\tilde x$ along the shortest geodesic connecting them, $L =
\gamma + \frac 12 \ln\frac{m^2|\sigma|}{2}$ and $\gamma$ is
Euler's constant. In the above expression we have to take all
terms which survive after taking two derivatives and going to the
coincidence limit. Here $a_k$ are the heat kernel coefficients
(see, for example \cite{Vas03}).

The spacetime under consideration (\ref{metric}) possesses the
singular curvature (\ref{ScaCur}) and therefore we have a
spacetime with the singular surface, $\rho =0$, with codimension
one. In this case we have to use expressions for heat kernel
coefficients obtained in Ref. \cite{GilKirVas01} (see also Ref.
\cite{Vas03}). The main result of the paper \cite{GilKirVas01} is
that the heat kernel coefficients in this case may be represented
as a sum of standard expressions for heat kernel coefficients
calculated without a singular surface and surface terms:
\begin{equation}
a_n (M,\Sigma) = a_n(M\setminus \Sigma) + a_n^{\Sigma}(\Sigma),
\end{equation}
where $M$ is a manifold, $\Sigma$ is a singular surface and
$a_0^{\Sigma}(\Sigma) =0$. The spacetime given by the metric
(\ref{metric}) is flat and therefore $a_0(M\setminus \Sigma) =1,\
a_{n>1}(M\setminus \Sigma)=0$. All surface terms
$a_n^{\Sigma}(\Sigma)$ are localized at the singular surface. For
example
\bs
\begin{eqnarray}
a_1^{\Sigma}(\Sigma) &=& -\left(\frac 16 - \xi \right)\frac 8a
\delta (\rho) = \left(\frac 16 - \xi \right) R,\adb\\
a_2^{\Sigma}(\Sigma) &=& \frac{256}{3a^3} \xi^3\delta (\rho).
\end{eqnarray}
\es
Therefore for $\rho\not = 0$ all singular contributions are zero
and we find
\begin{equation}\label{DS}
 G_{DS}(x;\tilde x) = \frac{1}{8\pi^2 \sigma} + \frac{m^2}{8\pi^2} \left[\gamma +
 \frac
12 \ln\frac{m^2|\sigma|}{2}\right] - \frac{m^2}{16\pi^2}.
\end{equation}
In fact, this expression for $G_{DS}(x;\tilde x)$ does not contain
any curvature terms and therefore coincides with that of Minkowski
space-time. Because of this fact the renormalization procedure
reduces to discarding the term $g^{M}_{\omega l}(\rho,\tilde\rho)$
in Eq. \Ref{gr}. Now, by using the formulas \Ref{GE}, \Ref{gr},
\Ref{phisq} and \Ref{DS}, we obtain
\begin{equation}\label{phisqseries}
\langle\phi^2(\rho)\rangle = -\frac{1}{2\pi^2 r}\int_0^\infty
d\omega\sum_{l=0}^\infty \nu \left.\frac{\lambda a(I_\nu K_\nu' +
I_\nu'K_\nu) + (8\xi-1) I_\nu K_\nu}{2\lambda a K_\nu K_\nu' +
(8\xi-1) K_\nu^2}\right|_{\lambda a} K^2_\nu(\lambda r).
\end{equation}

For $p=-(1+\zeta)/2 = 4\xi-1$ we observe singularity in integrand
for $l=0$ for $z^2 = (4\xi-1)^2 - m^2 a^2$. It is possible for
$\xi > 1/4$. For $\xi \le 1/4$ there is no singularity. We will
assume $\xi \le 1/4$.

Another singularity is for $\rho=0$. Indeed for $\rho=0$ the
uniform expansion of integrand gives
\begin{equation}
\langle\phi^2(\rho)\rangle = -\frac{1}{2\pi^2 a^2} \int_0^\infty d
z \sum_{l=0}^\infty \left\{\frac 14 t^2(1+\zeta - t^2) - \frac
1{8\nu} t^3(1+\zeta - t^2)^2  + O(\nu^{-2})\right\},
\end{equation}
with $t=1/\sqrt{1+z^2}$. The first two terms of the series are
divergent.

\begin{figure}[ht]
\begin{center}
{\epsfxsize=7truecm\epsfbox{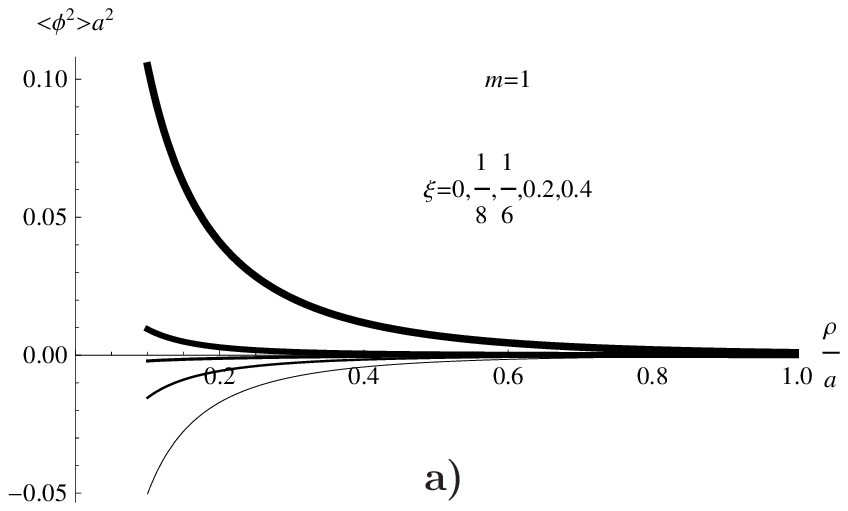}}\hspace{2em}
{\epsfxsize=7truecm\epsfbox{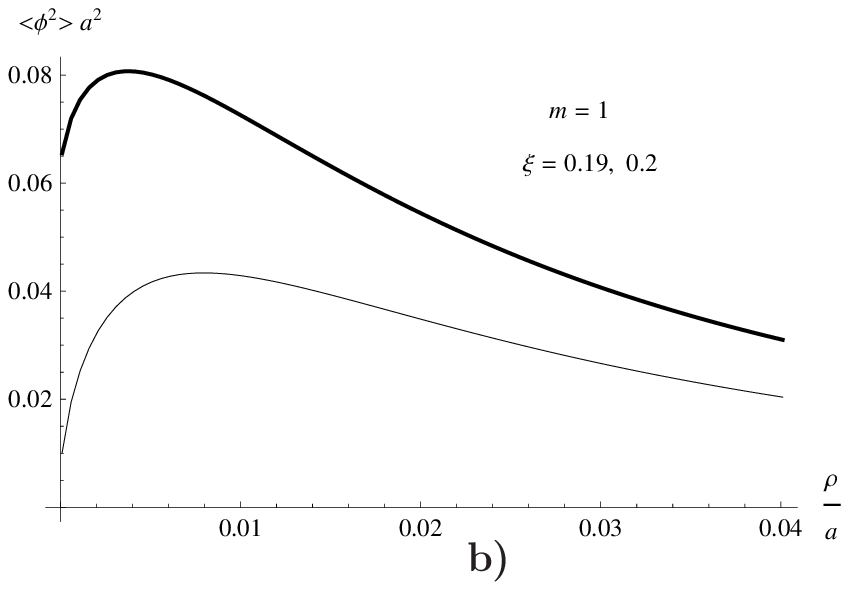}}
\end{center}\caption{Plots of $\phisq\,a^2$ for $\xi =
0,\frac 18, \frac 16, 0.2, 0.4$ from bottom to up. The value of
the field mass $m$ is fixed: $m=1$. The $\phisq\,a^2$ behaviour  is shown on the figure $b$ for small distances at the throat. All lines fall down to minus infinity near the throat. }\label{fig:1ab}
\end{figure}

\begin{figure}[ht]
\begin{center}
{\epsfxsize=7truecm\epsfbox{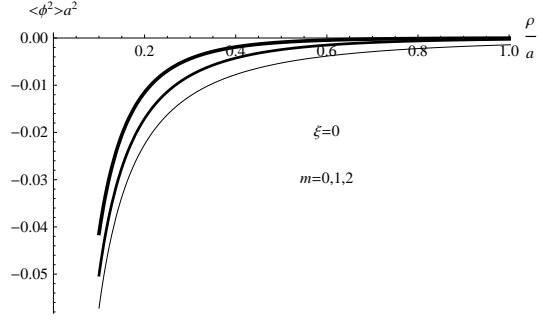}}
\end{center}\caption{Plots of $\phisq\,a^2$ for $m=0,1,2$ from bottom to up. The value of
$\xi$ is fixed: $\xi =0$. \label{fig:2}}
\end{figure}

The result of numerical analysis of $\phisq$ given by Eq.
\Ref{phisqseries} is shown in Figs. \ref{fig:1ab} and
\ref{fig:2}.

\section{Vacuum stress-energy tensor $\langle T_{\mu\nu}\rangle$}\label{Sec5}
To calculate the stress-energy tensor one may use the standard
formula
\begin{eqnarray}
 \langle T_{\mu\nu}\rangle^{ren} &=& \lim_{x'\to x}\left\{ \left(\frac 12
-\xi\right)(g_\mu^{\ \alpha'} G_{;\alpha' \nu} + g_\nu^{\
\alpha'}G_{;\mu\alpha'}) + \left(2\xi - \frac
12 \right) g_{\mu\nu}g^{\sigma\alpha'} G_{;\sigma\alpha'} - \xi (G_{;\mu\nu} +
g_\mu^{\ \alpha'} g_\nu^{\ \beta'} G_{;\alpha'\beta'})\right.\adb\nonumber\\
&+&\left. 2\xi g_{\mu\nu} (m^2 + \xi R)G + \xi \left( R_{\mu\nu} - \frac 12
g_{\mu\nu}R\right) G - \frac 12 m^2 g_{\mu\nu}G\right\},
\end{eqnarray}
where $G$ is the renormalized Euclidean Green function. For above
function it is easy to show that
\begin{eqnarray*}
 [G_{,\mu}] &=& [G_{,\mu'}],\adb\\{}
 [G_{,\mu}] &=& 0,\ \mu \not = \rho \adb\\{}
 [G_{,\mu\nu}] &=& [G_{,\mu'\nu'}],\adb\\{}
 [G_{;\mu\nu}] &=& [G_{;\mu'\nu'}],\adb\\{}
 [G_{,\mu\nu}] &=& - [G_{,\mu\nu'}], \ \mu,\nu \not = \rho.
\end{eqnarray*}
For this reason we may rewrite the expression for EMT in the following form
\begin{eqnarray}
 \langle T_{\mu\nu}\rangle &=& \left[(1-2\xi)G_{;\mu\nu'} -
2\xi G_{;\mu\nu} + \left(2\xi - \frac 12\right) g_{\mu\nu}
(G_{;\alpha}^{;\alpha'} + (m^2 + \xi R) G) + \xi R_{\mu\nu} G\right].
\end{eqnarray}
For calculation the EMT we have to take into account that $P_l(1)
=0$ and $P'_l(1) = l(l+1)/2$. For brevity let us define operator
$\hat{\cal L}$ and the function $f$
\begin{eqnarray}
 \hat{\cal L}\{\cdot\} &=& -\frac{1}{2\pi^2}\int_0^\infty d\omega\sum_{l=0}^\infty
\nu \left.\frac{\lambda a(I_\nu K_\nu' + I_\nu'K_\nu) +
(8\xi-1) I_\nu K_\nu}{2\lambda a K_\nu K_\nu' + (8\xi-1)
K_\nu^2}\right|_{\lambda a} \{\cdot \}\adb\\
f &=& \frac{K_\nu (\lambda r)}{\sqrt{r}}.
\end{eqnarray}
The components of the EMT have the following form
\begin{eqnarray}
    \langle T_{t}^{t}\rangle &=& \hat{\cal L}\left\{\omega^2f^2  +
\left(2\xi - \frac 12\right) \left(\left(\lambda^2 + \xi R +
\frac{l(l+1)}{r^2}\right)f^2 + \dot f^2\right)\right\},\adb \label{Ttt} \\
    \langle T_{\rho}^{\rho}\rangle &=& \hat{\cal L}\left\{(1-2\xi)\dot f^2 -2\xi
\ddot f f + \left(2\xi - \frac 12\right) \left(\left(\lambda^2 + \xi R +
\frac{l(l+1)}{r^2}\right)f^2 + \dot f^2\right) + \xi R^\rho_\rho
f^2\right\},\adb \label{Trr}\\
    \langle T_{\theta}^{\theta}\rangle &=& \hat{\cal L}\left\{\frac{l(l+1)}{2r^2}
f^2  - \frac{2\xi}{r} \dot f f + \left(2\xi - \frac 12\right)
\left(\left(\lambda^2 + \xi R +
\frac{l(l+1)}{r^2}\right)f^2 + \dot f^2\right) + \xi R^\theta_\theta
f^2\right\},\adb\\
\langle T_{\varphi}^{\varphi}\rangle &=& \langle
T_{\theta}^{\theta}\rangle, \label{Tthth}
\end{eqnarray}
where
an overdot notes the derivative with respect to $\rho$. The
function $f$ obeys to the equation
\begin{equation*}
 \ddot f + \frac{2}{r}\dot f - \left(\frac{l(l+1)}{r^2} +
\lambda^2\right)f = 0.
\end{equation*}

\section{Numerical analysis}\label{Sec6}
The expressions (\ref{Ttt})--(\ref{Tthth}) for components of
$\Tmn$ are not much suitable for an analytical consideration,
therefore, we have applied numerical methods for their analysis.
In this section we will discuss results of numerical analysis.

For a static spherically symmetric configuration one has $\langle
T^{t}_{t}\rangle=-\varepsilon$, $\langle T^\rho_\rho\rangle=p$,
and $\langle T^\theta_\theta\rangle=\langle
T^\varphi_\varphi\rangle=p_{t}$, where $\varepsilon$ is the energy
density, $p$ is the radial pressure, and $p_t$ is the transverse
pressure. Values of $\varepsilon$, $p$, and $p_t$ are connected by
the conservation law $\langle T^\mu_\nu\rangle_{;\mu}=0$, which
for the Minkowsky metric \Ref{Mink} takes the simple form:
\beq
p_t=p+{\textstyle\frac12}rp'.
\eeq
From here one may easily find $p_t$ provided $p$ is found.

\begin{figure}[ht]
\begin{center}
{\epsfxsize=8truecm\epsfbox{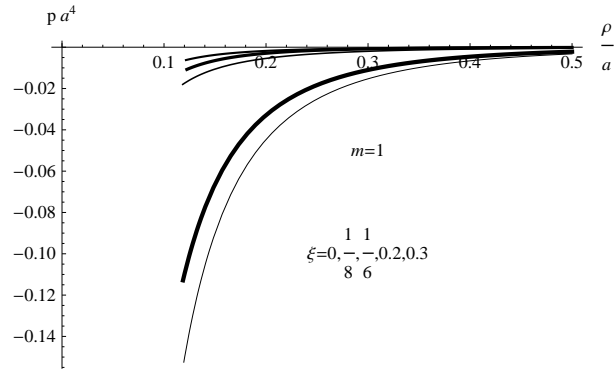}}\vspace{5em}
\end{center}\vspace*{-6em}
\caption{Plots of $p\, a^4$ for $m=1$ and $\xi = 0,\frac 18, \frac
16, 0.2, 0.3$. The thicker is the line, the grater is the value of
$\xi$. }\label{fig:3}
\end{figure}
\begin{figure}[ht]
\begin{center}
{\epsfxsize=8truecm\epsfbox{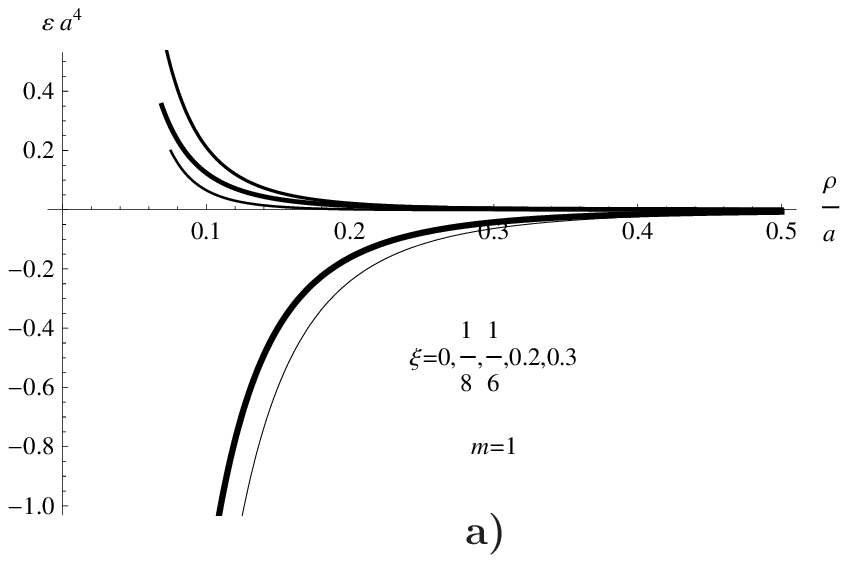}}%
{\epsfxsize=8truecm\epsfbox{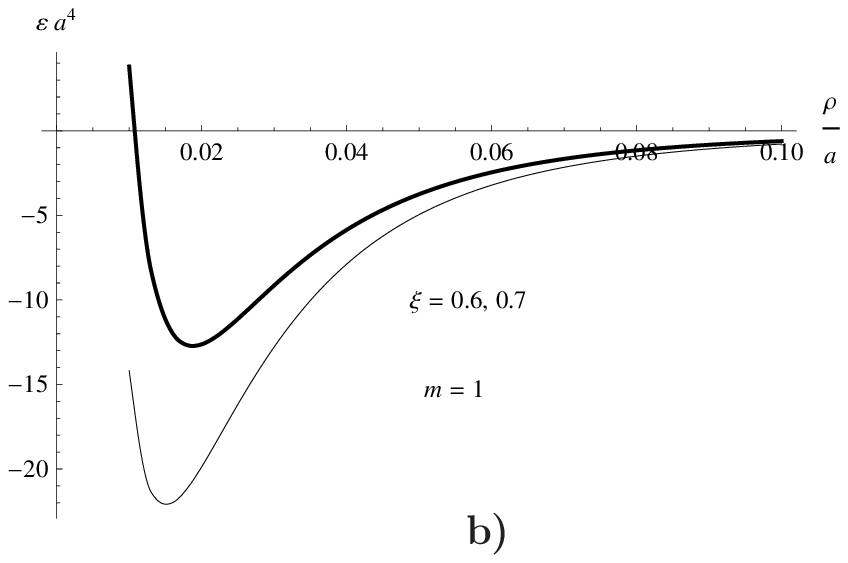}}
\end{center}
\caption{Plots of $\varepsilon a^4$ for $m=1$ and $\xi = 0,\frac
18, \frac 16, 0.2, 0.3$.  The thicker is the line, the grater is
the value of $\xi$. The $\varepsilon a^4$ behaviour  is shown on the figure $b$ for small distances at the throat. All lines fall down to infinity near the throat.}\label{fig:4ab}
\end{figure}
\begin{figure}[ht]
\begin{center}
{\epsfxsize=8truecm\epsfbox{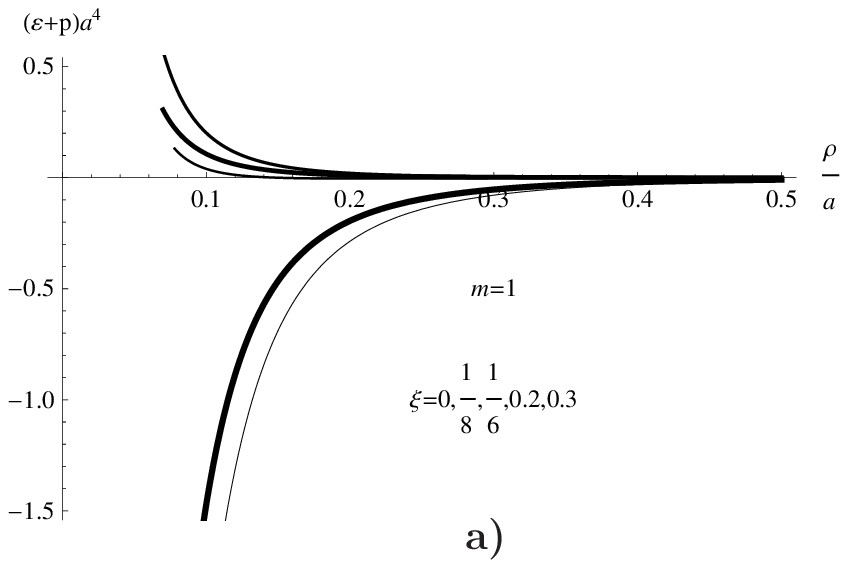}}%
{\epsfxsize=8truecm\epsfbox{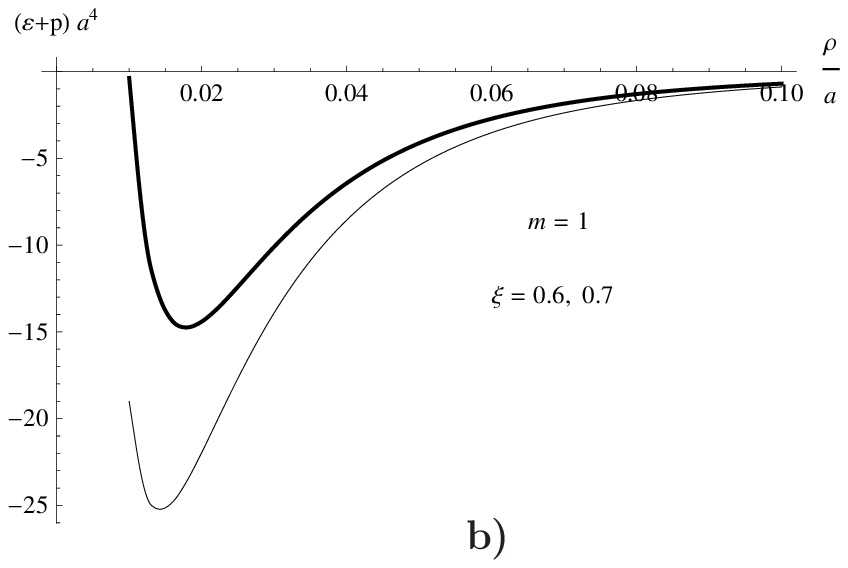}}
\end{center}
\caption{Plots of $(\varepsilon +p)a^4$ for $m=1$ and $\xi =
0,\frac 18, \frac 16, 0.2, 0.3$.  The thicker is the line, the
grater is the value of $\xi$. The $(\varepsilon +p)a^4$ behaviour  is shown on the figure $b$ for small distances at the throat. All lines fall down to infinity near the throat.}\label{fig:5ab}
\end{figure}

We compute numerically the energy density $\varepsilon=-\langle
T^{t}_{t}\rangle$ and the radial pressure $p=\langle
T^{\rho}_{\rho}\rangle$ using Eqs. \Ref{Ttt}, \Ref{Trr}. Results
of numerical computations are given in Figs.
\ref{fig:3}--\ref{fig:5ab}. Let us discuss them in details. In Fig.
\ref{fig:3} plots of $p(\rho)$ are shown for a fixed value of $m$
and various values of the curvature coupling parameter $\xi$. It
is seen that $p$ is everywhere negative, and $p\to-\infty$ in the
limit $\rho\to0$. Thus, a vacuum polarization leads to an
infinitely negative radial pressure $p$ at the wormhole's throat.

Plots of energy density $\varepsilon(\rho)$ are shown in Fig.
\ref{fig:4ab}.
It is seen that a qualitative behavior of $\varepsilon$ depends on
$\xi$. Provided $\xi<1/8$ or $\xi>0.2$, the function
$\varepsilon(\rho)$ reaches a negative minimum at some $\rho$, and
tends to zero far from the throat. In case $1/8<\xi<0.2$
$\varepsilon$ is everywhere positive monotonically decreasing
function. It is worth noting that in both cases the vacuum energy
density goes to infinity at the wormhole's throat, i.e.,
$\varepsilon\to\infty$ in the limit $\rho\to0$ (see Fig.
\ref{fig:4ab}b).

It is particularly important for a wormhole geometry to check
whether the vacuum stress energy tensor obeys the usual energy
conditions. In particular, the null energy condition (NEC) reads
$\langle T_{\mu\nu}\rangle k^\mu k^\nu\ge 0$, where $k^\mu$ is an
arbitrary null vector. In a static spherically symmetric case the
NEC reduces to $\varepsilon+p\ge 0$. Graphs for the combination
$\varepsilon+p$ are given in Fig. \ref{fig:5ab}. It is seen that
the combination $\varepsilon+p$ behaves similarly to the energy
density $\varepsilon$. In particular, $\varepsilon+p\to\infty$ in
the limit $\rho\to0$ (see Fig. \ref{fig:5ab}b). Thus, the vacuum
stress energy tensor $\langle T_{\mu\nu}\rangle$ does not violate
the NEC in the vicinity of the wormhole's throat.

\section{Conclusion}
We have calculated the vacuum polarization $\langle\phi^2\rangle$
and components of the vacuum stress energy tensor $\langle
T_{\mu\nu}\rangle$ of the massive scalar field in a wormhole
spacetime using the short-throat flat-space approximation for the
wormhole geometry. The most important result obtained consists in
the fact that $\langle T_{\mu\nu}\rangle$ does not violate the NEC
in the vicinity of the wormhole's throat. As a consequence, this
implies that the vacuum polarization cannot self-consistently
support the wormhole. Of course, it is necessary to emphasize that
this conclusion has been obtained for the very simple model of
wormhole. To make more general conclusions one should consider
more realistic models of wormholes.

\subsection*{Acknowledgments}
This work was supported in part by the Russian Foundation for
Basic Research grants No. 08-02-00325.


\end{document}